\documentclass{iau}
\usepackage{graphicx}

\title[Small and large dust grains in transitional disks] 
{Can a planet explain different cavity sizes for small \& large dust grains in transition disks?}

\author[A.\ Garufi, H.\ Avenhaus, S.\,P.\ Quanz]   
{Antonio Garufi$^1$,
  Henning Avenhaus$^1$,
  Sascha P. Quanz$^1$}

\affiliation{$^1$Institute for Astronomy, ETH Zurich, Wolfgang-Pauli-Strasse 27, Zurich, Switzerland \\ email: {\tt antonio.garufi@phys.ethz.ch} }

\pubyear{2013}
\volume{299}  
\pagerange{--}
\jname{Exploring the formation and evolution of planetary systems}
\editors{Brenda Matthews \& James Graham}
\begin{document}

\maketitle

\begin{abstract}
Dissimilarities in the spatial distribution of small ($\mu$m$-$size) and large (mm$-$size) dust grains at the cavity edge of transition disks have been recently pointed out and are now under debate. We obtained VLT/NACO near-IR polarimetric observations of SAO 206462 (${\rm HD 135344B}$). The disk around the star shows very complex structures, such as dips and spirals. We also find an inner cavity much smaller than what inferred from sub-mm images. The interaction between disk and orbiting companion(s) may explain this discrepancy.

\end{abstract}

\firstsection 
\section{Introduction}
A small sample of disks, the transition disks, shows a peculiar dip at infrared wavelengths, suggesting a depletion of warm dust around the central star (\cite[Strom et al. 1989]{Strom1989}). Disk$-$companion interaction (\cite[Rice et al. 2003]{}), photoevaporation (\cite[Alexander \& Armitage 2007]{}), and particle growth (\cite[Dullemond \& Dominik 2005]{}) are possible clearing processes.

Polarimetric Differential Imaging (PDI) is allowing high-resolution imaging of circumstellar disks (e.g. \cite[Quanz et al. 2011]{}, \cite[Hashimoto et al. 2012]{}) with unprecedented inner working angle ($0.1''$). Recently, comparisons of PDI images with sub-millimeter images (e.g. \cite[Andrews et al. 2011]{}) have revealed different spatial distribution for small and large dust grains (see e.g.\ \cite[Dong et al. 2012]{}). 

\begin{figure}[h]
\begin{center}
 \includegraphics[width=1.4in]{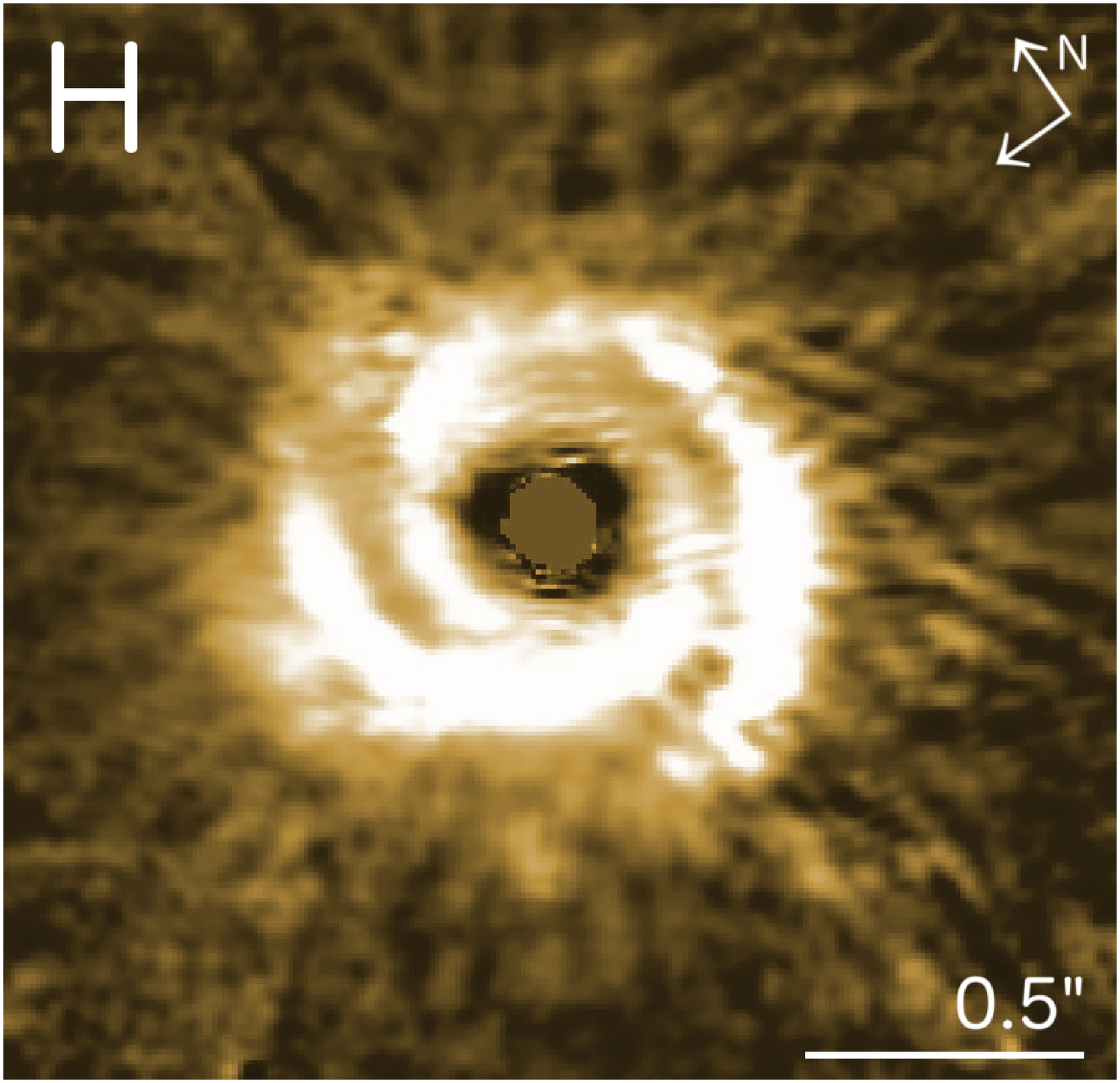}
  \includegraphics[width=1.4in]{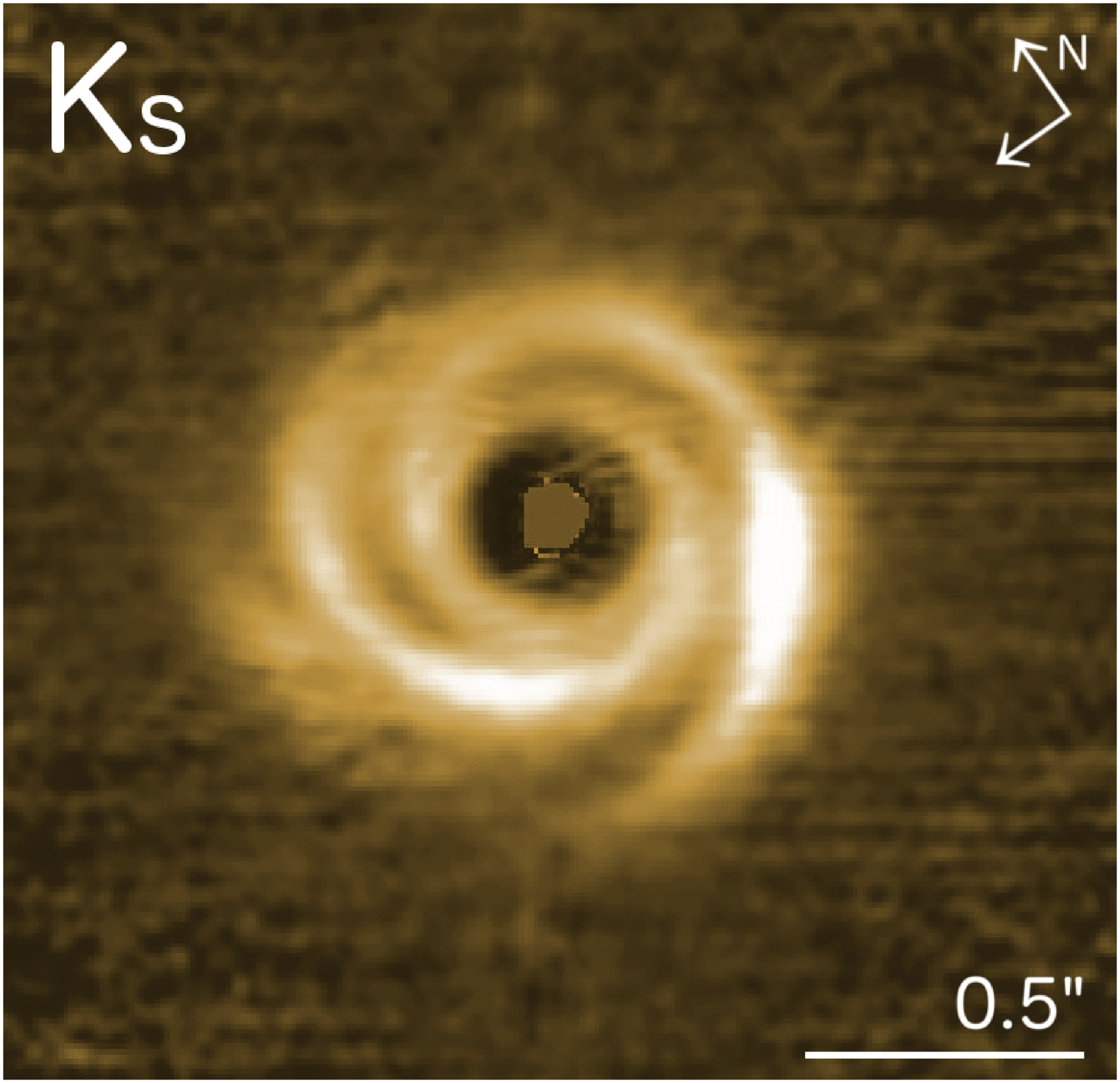} 
 \caption{The disk around SAO 206462 from PDI VLT/NACO observations.}
  \label{images}
\end{center}
\end{figure}
\section{PDI observations and interpretation}

PDI observations of the Herbig Ae/Be SAO 206462 (HD 135344B) were obtained with the high-resolution NACO (\cite[Lenzen et al. 2003]{}, \cite[Rousset et al. 2003]{}) at the Very Large Telescope (VLT), in $H$ and $K_{\rm S}$ band. The basic principle of PDI is the simultaneous imaging of the linear polarization of the source along two orthogonal directions. The detailed observation setting and data reduction can be found in \cite[Quanz et al. (2011)]{} and Avenhaus et al. (in prep.). The final products are radial Stokes $Q_{r}$ parameter images of the source described in Garufi et al. (in prep.) and shown in Fig.\,\ref{images}. The disk is revealed in scattered light in both bands. The images show three main peculiarities: an inner cavity (inside ${28 \pm 6 {\rm \ AU}}$) with light depleted by a factor down a few tenths, a quasi-circular rim surrounding the cavity, and two spiral arms extending from the rim outward. 

Apart from the spiral structure (probably due to a companion orbiting at large scale, \cite[Muto et al. 2012]{Muto2012}), the most tantalizing aspect suggested by these images is the different cavity size of small grains (28 AU) with respect to large grains (39 AU, \cite[Brown et al. 2009]{Brown2009}). Similar discrepancies were recently pointed-out by \cite[Dong et al. (2012)]{Dong2012}. 

The cavity for small grains is likely to be due to tidal interaction with a companion. Photoevaporation and dust grain growth are indeed ruled-out: because of high accretion rate (\cite[Sitko et al. 2012]{Sitko2012}) and sub-AU inner dust belt (\cite[Fedele et al. 2008]{Fedele2008}) the former, abrupt radial profile (this work) and absence of mm-size grains (\cite[Brown et al. 2009]{Brown2009}) the latter. 

\begin{figure}[h]
\begin{center}
 \includegraphics[width=3.8in]{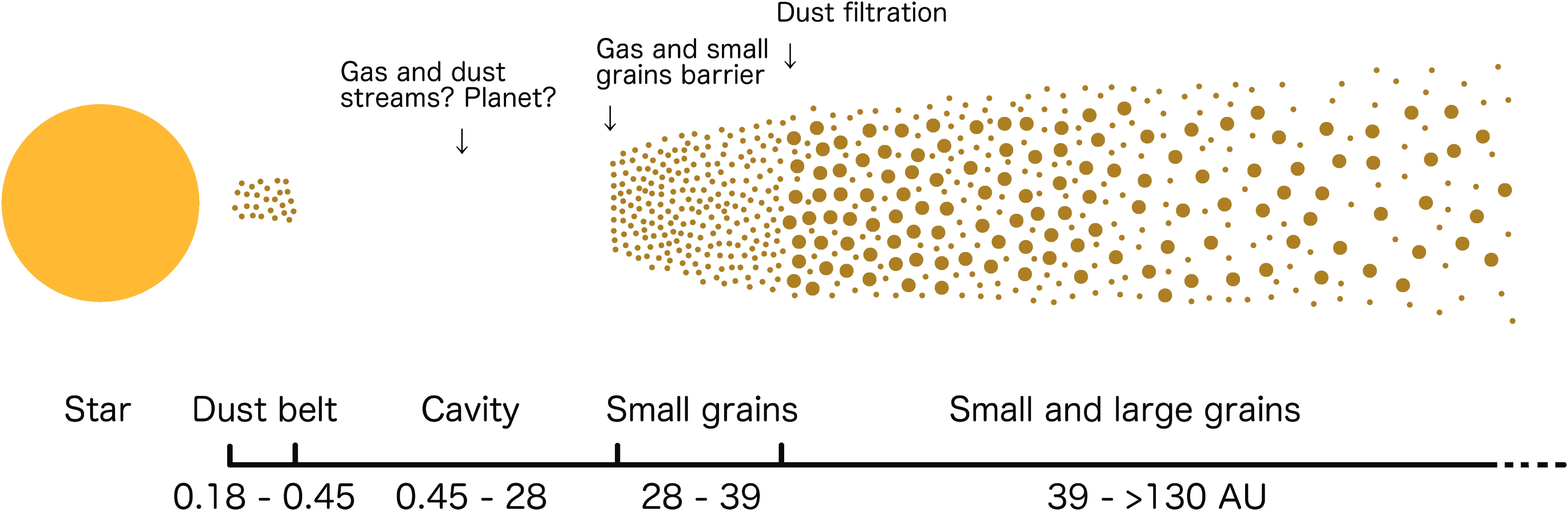}
 \caption{Illustrative sketch showing the dust radial distribution in the disk of SAO 206462}
  \label{sketch}
\end{center}
\end{figure}

The scenario with clearing by a companion orbiting inside the cavity may also explain the observed dissimilarity in the cavity sizes. In fact, the pile-up of large grains due to a giant planet can occur at up to 10 tidal radii (\cite[Pinilla et al. 2012]{Pinilla2012}), whereas the outer edge of the gaseous halo cannot exceed 5 tidal radii (\cite[Dodson-Robinson \& Salyk 2011]{Dodson-Robinson2011}). We suggest a scenario (see Fig.\,\ref{sketch}) where a giant planet is generating a pressure bump at 39 AU, which holds back mm-size grains but allows $\mu$m-size grains to be dragged inward along with the gas as down as 28 AU. Non-keplerian flows of gas and small dust grains can still be present in the cavity, thus to sustain the inner dust belt and the high accretion rate of the source. We analytically find that a 5 to 15 $\rm M_J$ at 17 to 20 AU is consistent with the observed cavity sizes. However, a multiple-planets system is not ruled-out.   

\end{document}